\documentclass[a4paper,11pt]{article}
\pdfoutput=1 

\usepackage{jinstpub}
\usepackage{url,lineno,subcaption}
\usepackage{microtype}
\usepackage{braket}
\usepackage{textcomp}
\usepackage{amssymb}
\usepackage{graphicx}

\title{Signal formation and sharing in AC-LGADs using the ALTIROC 0 front-end chip}

\author[a]{G.~D'Amen\footnote{corresponding author},}
\author[b]{W.~Chen,} 
\author[c]{C.~De~La~Taille,} 
\author[b]{G.~Giacomini,}
\author[d]{D.~Marchand,}
\author[c]{M.~Morenas,}
\author[d]{C.~Munoz~Camacho,}
\author[b]{E.~Rossi,}
\author[c]{N.~Seguin-Moreau,}
\author[d]{L.~Serin,}
\author[a]{A.~Tricoli,}
\author[d]{P.-K.~Wang}

\affiliation[a]{Brookhaven National Lab, Physics Department, Upton (NY), 11973, US}
\affiliation[b]{Brookhaven National Lab, Instrumentation Division, Upton (NY), 11973, US}
\affiliation[c]{OMEGA-CNRS-Ecole Polytechnique, Route de Saclay, 91128 Palaiseau, France}
\affiliation[d]{Universite Paris-Saclay, CNRS/IN2P3, IJCLab, Orsay, France}

\emailAdd{gdamen@bnl.gov}

\abstract{
    The development of detectors that provide high resolution in four dimensions has attracted wide-spread interest in the scientific community for applications in high-energy physics, nuclear physics, medical imaging, mass spectroscopy as well as quantum information. However, finding a technology capable of fulfilling such aspiration proved to be an arduous task.
    Among other silicon-based candidates, the Low-Gain Avalanche Diode (LGAD) has already shown excellent timing performances but proved to be unsuitable for fine pixelization. Therefore, the AC-coupled LGAD (AC-LGAD) approach was introduced to provide high resolution in both time and space, making it a promising candidate for future 4D detectors. 
    However, appropriate readout electronics must be developed to match the sensor's fast-time and fine-pitch capabilities. This is currently a major technological challenge.
    In this paper, we test AC-LGAD prototypes read out by the fast-time ASIC ALTIROC 0, originally developed for the readout of DC-coupled LGADs for the ATLAS experiment at the HL-LHC. Signal generated by either betas from a $^{90}$Sr source or a focused infra-red laser were analyzed. This paper details the first successful readout of an AC-LGAD sensor using a readout chip. This result will pave the way for the design and construction of a new generation of AC-LGAD-based 4D detectors.
}

\begin{document}

\maketitle
\flushbottom

\section{Introduction}
Low-Gain Avalanche Diodes (LGADs) are silicon sensors engineered for the fast detection of minimum ionizing particles (mips)~\cite{Fern_ndez_Mart_nez_2016}. Being built on thin silicon substrates (order of 20 - 50~$\mu$m) and thanks to the presence of a gain layer that boosts the signal by a factor of a few tens, LGADs can achieve a timing resolution below 30~ps~\cite{Cartiglia_2017}. LGADs were prototyped by CNM (Barcelona, Spain) under the auspices of the RD50 collaboration at CERN. Nowadays, LGADs sensors are successfully designed and fabricated by several facilities around the world. Since their introduction, the interests in the High-Energy Physics (HEP) community has grown, as detectors based on LGADs will constitute the High Granularity Timing Detector (HGTD)~\cite{HGTD} and MIP Timing Detector (MTD)~\cite{MTD} upgrades of the ATLAS and CMS experiments at CERN (Geneva, Switzerland), respectively.
\begin{figure}[ht]
    \centering
    \includegraphics[width=\textwidth]{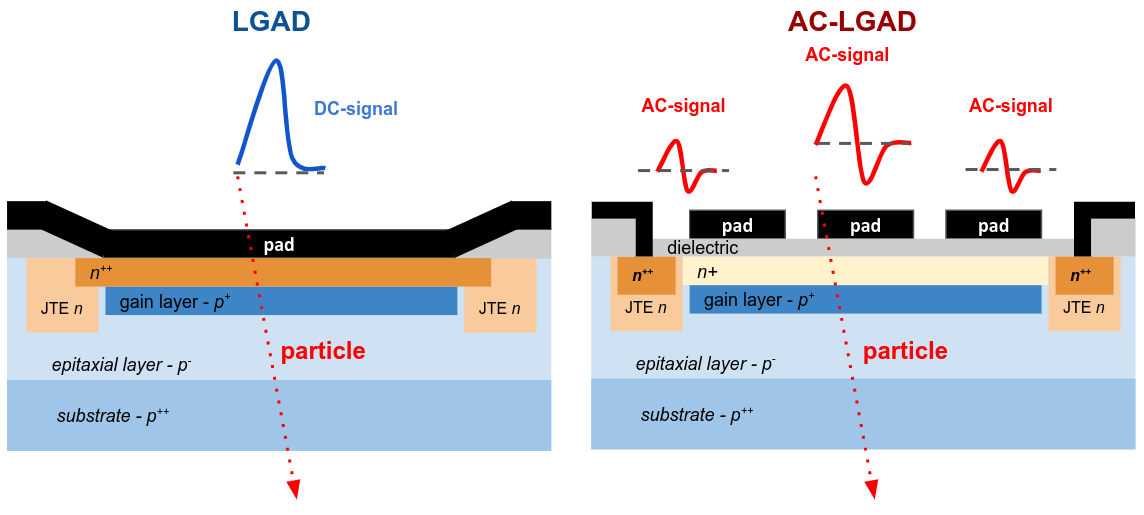}
    \caption{Signal formation in LGAD (\textit{left}) and AC-LGAD (\textit{right}) sensors. In AC-LGADs, signals are capacitively induced to metal electrodes separated from the active area by a thin dielectric film. The AC-coupled signal is shared among multiple electrodes.}
    \label{fig:lgad_vs_aclgad}
\end{figure}
However, it has been observed that mips crossing close by the pad edge in a LGAD sensor generate signals that are not amplified, as the generated electrons - in their  drift to the n-type pixel - do not go through the high electric field. To achieve a uniform multiplication in most of the area and a high fill factor, the pad pitch must by far greater than the substrate thickness. Pixel or strip pitches in the order of $\sim$100~$\mu$m are therefore not achievable in LGAD technology~\cite{Andra:ay5534}.

To overcome this severe limitation, the AC-LGAD concept was developed~\cite{aclgad_1, aclgad_2, aclgad_3, aclgad_4}, in which the signal is capacitively induced to metal electrodes placed over the active area but separated from it by a thin dielectric film made of silicon oxide or nitride, among others (Fig.~\ref{fig:lgad_vs_aclgad}). Since their fabrication process is similar to that for standard LGADs, many LGAD facilities around the world started developing AC-LGADs too. Fabrications aiming at process and design parameters optimization are on-going, as well as intense testing campaigns. One of the consequences of the AC-coupling of signal in AC-LGADs is that signal is shared among neighboring electrodes. Such a property was demonstrated to provide precise spatial resolution with sparse electrodes~\cite{Apresyan_2020}. 

The most significant limitation for the broad use of AC-LGADs in scientific applications is the lack of readout electronics that match the precise spatial and time resolution of AC-LGADs. This article demonstrates that AC-LGADs can be read out by a state-of-the-art ASIC developed for the acquisition of signals from standard DC-coupled LGADs.

\subsection{AC-LGAD sensors as 4D detectors}

In recent beam tests, AC-LGADs have demonstrated to achieve spatial resolution better than one tenth of the strip pitch and timing resolution compatible with DC-coupled LGADs~\cite{Heller_2022}. Such good spatial resolution results from the interpolation of signals shared among several AC-coupled electrodes. While this signal-sharing feature may lead to higher readout occupancy, especially in high-multiplicity environment, it also makes AC-LGADs first-class candidates for 4D detectors.

The characterization studies of LGAD and AC-LGAD sensors have been performed so far on custom-made test-boards (see for example Refs.~\cite{Giacomini_2019_LGAD}\cite{Giacomini_2019_ACLGAD}), designed for the evaluation of a small quantity of sensors. These readout test-boards are based on discrete electronics, by exploiting either low-noise fast trans-impedance or RF amplifiers that closely reproduce the current pulse generated by the sensor. However, these boards are not optimized for power dissipation and can only readout a limited number of channels; in addition, they typically do not provide on-line analysis capabilities without the aid of an external fast oscilloscope.

When looking at large-scale, multi-channel systems targeting 4D tracking, it becomes clear the need to evaluate the AC-LGAD performance when coupled to higher-complexity readout systems. For example, in the EIC (Electron Ion Collider) experiments, AC-LGADs are the sensor baseline to be used in the Roman Pots detector~\cite{romanpots}. The pixel size is expected to be 0.5 mm$\times$0.5 mm, almost 7 time smaller than that used in HGTD/MTD. It is foreseen that the Roman Pots detector at the EIC will be based on a readout chip tailored to match the features of AC-LGAD signals and will be based upon an existing readout ASIC designed for fast-timing and high-bandwidth particle physics experiments. Such ASIC, called the ATLAS LGAD Timing Integrated ReadOut Chip (ALTIROC), is demonstrated here to preserve the signal characteristics that are typical of AC-LGADs, for example signal sharing, and allow fast and precise tracking.

The ALTIROC chip was developed~\cite{delataille:hal-02058308} to readout DC-coupled signals generated in LGAD sensors while satisfying the strict timing requirements of the upcoming ATLAS HGTD foreseen for the 2028 High-Luminosity LHC upgrade. The ALTIROC chip was used to characterize the response of an AC-LGAD sensor to either betas from a $^{90}$Sr source or a focused infra-red (IR) laser. Results obtained using this setup were compared to the characterization of an identical AC-LGAD strip sensor attained using a 16-channels readout board designed and produced at the Fermi National Accelerator Laboratory (FNAL, Batavia, IL, USA).

Section~\ref{sec:Setup} details the experimental setup used in this study, with particular attention to the layout of the characterized AC-LGAD sensor, the readout capabilities of the ALTIROC chip, and a breakdown of the particle sources employed in the study. 

Section~\ref{sec:signal_characterization} presents the response of the AC-LGAD sensor readout by the ALTIROC chip to either an uncollimated flux of beta particles and a focused IR laser. This is followed by a detailed evaluation of the generated signal amplitude, full-width at half-maximum (FWHM), jitter and signal sharing between strips. Finally, the differences in response and signal shape with an identical AC-LGAD sensor read out by a custom-made RF-amplifier based system are also presented.

Finally in Section~\ref{sec:digital_characterization} this response is compared to that of the fixed-threshold discriminator integrated into the ALTIROC chip, allowing for the measurement of the time over threshold (ToT) of the particle signal. The ToT enables the use of center-of-gravity algorithms to reconstruct the particle position on the sensor with sub-pitch precision, thanks to the signal sharing capabilities of AC-LGADs.
\section{Experimental setup}\label{sec:Setup}
\begin{figure}[htbp]
    \textit{a)}\hspace{.49\textwidth}\textit{b)}
    \centering
    \includegraphics[width=.49\textwidth]{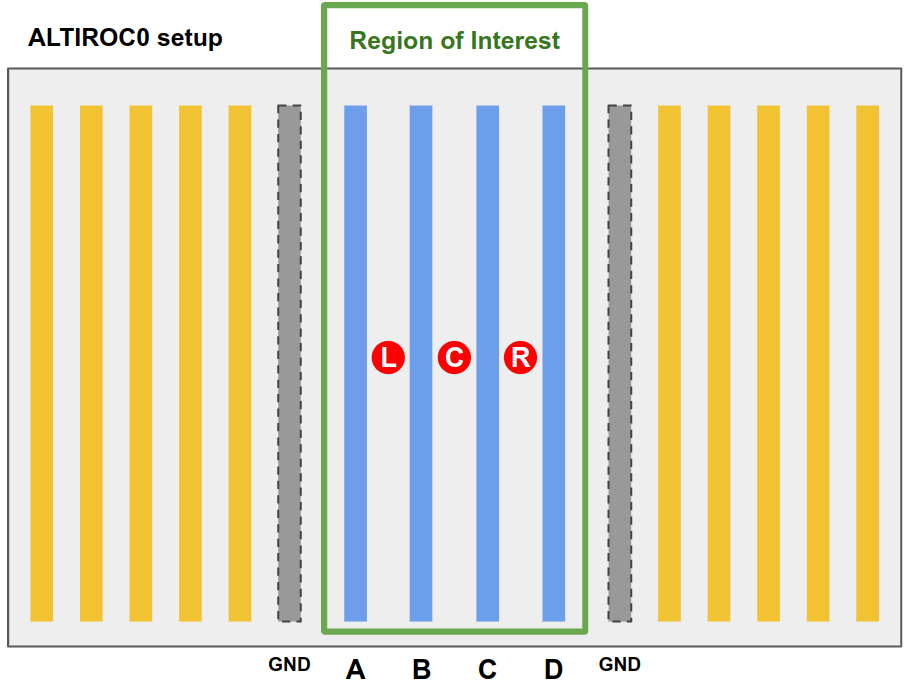}
    \includegraphics[width=.49\textwidth, trim = 2 0 0 0,clip]{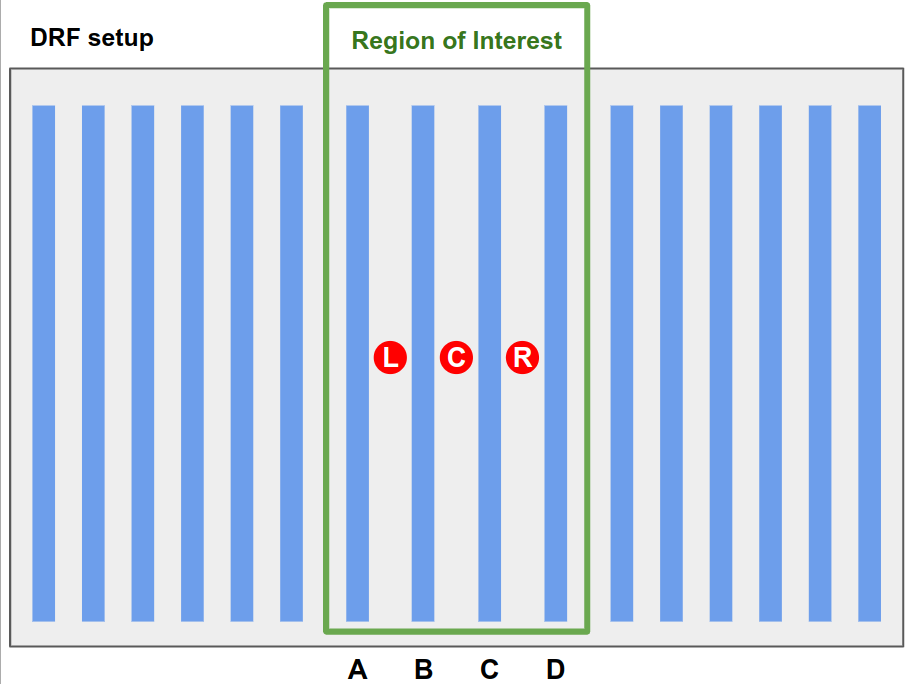}
    \caption{Schematic representation of the four AC-LGAD strips (strip A, B, C and D) wire-bonded to \textit{a)} ALTIROC 0 and \textit{b)} the DRF board. Strips in blue are wire-bonded to the readout systems while the ones in yellow are left floating. In \textit{a)}, the strips in grey are wire-bonded to the same ground (GND) as the ASIC to minimize the portion of the device that remains floating. In \textit{b)} all strips are wire-bonded to the readout system but only A, B, C and D are readout. The red points L, C and R are focusing points for the IR laser used to characterize the response of the detector.}
    \label{fig:wirebonding_scheme}
\end{figure}

The device under test is a strip AC-LGAD sensor fabricated at BNL. The sensor has an active area of 2$\times$2~mm$^2$, with the metallization layer divided in 16 strips with a 100~$\mu$m pitch separated by a gap 44~$\mu$m wide.
\begin{figure}[ht]
    \centering
    \includegraphics[width=.648\textwidth]{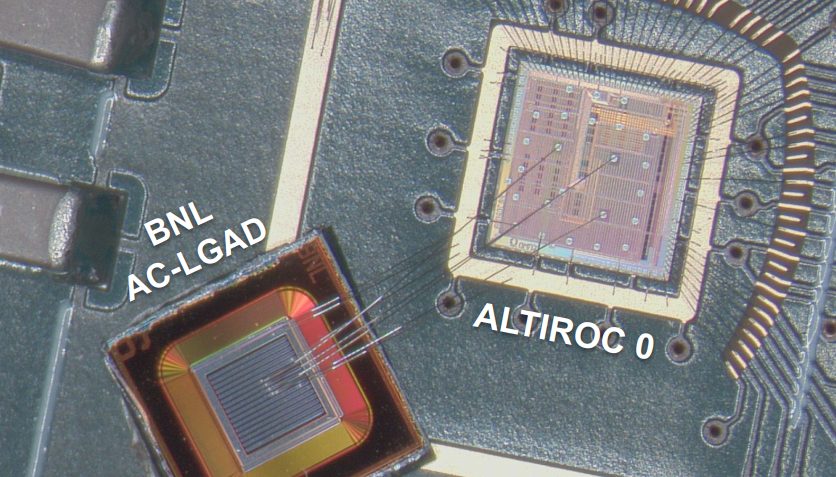}
    \includegraphics[width=.34\textwidth]{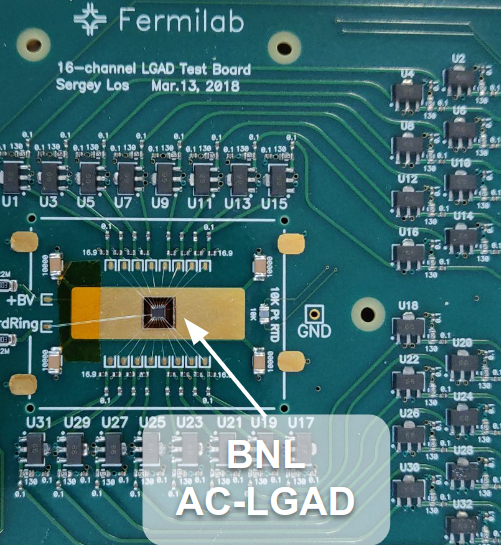}
    \caption{BNL AC-LGAD strip sensors wire-bonded to the ALTIROC 0 ASIC (\textit{left}) and a Discrete RF board (\textit{right}). Only the four central AC-LGAD strips are wirebonded in the ALTIROC 0 setup. All strips are connected to amplifiers in the Discrete RF board setup.}
    \label{fig:bondedAlti}
\end{figure}
Four neighbouring strips indicated as Strip A, Strip B, Strip C and Strip D, chosen to be far from the the device guard-ring in order to minimize border effects, were wire-bonded to the four Voltage Pre-Amplifier (VPA) channels on a ALTIROC chip of version 0-2B (referred to as ALTIROC 0 in the following), as shown in Fig.~\ref{fig:bondedAlti}. The strips on the left and right to this set were wire-bonded to the same ground as the ASIC to minimize the portion of the device that remains floating.  Fig.~\ref{fig:wirebonding_scheme} details the bonding scheme of the strips on the AC-LGAD setup. 
A bias voltage of V$_b$~=~-190~V was applied to the AC-LGAD via a Keithley 2410 source-meter to fully deplete the sensor. The ALTIROC 0 is wire-bonded to a custom testing board, mounting an ALTERA Cyclone-III FPGA for direct operation control and management of the data taking, and powered using an Agilent E3631A power supply.

This setup was compared to a twin AC-LGAD strip sensor, wire-bonded to the 16-channel readout board designed at FNAL seen in Fig.~\ref{fig:bondedAlti}. This test-board uses discrete RF components to translate current signals from the AC-LGAD into voltage. All sixteen strips of the AC-LGAD were wire-bonded to the input pads of this Discrete RF (DRF) board (Fig.~\ref{fig:wirebonding_scheme}). Each channel is equipped with a 2-stage amplifier chain as described in~\cite{Apresyan_2020}. 
Signals generated in the AC-LGAD by interactions with either betas or the IR laser were acquired using a LeCroy Waverunner 9404M-MS Oscilloscope\footnote{https://teledynelecroy.com/oscilloscope/waverunner-9000-oscilloscopes/waverunner-9404m-ms} using the high-sampling speed channels.
\subsection{The ALTIROC 0 chip}
\begin{figure}[ht]
    \centering
    \includegraphics[width=.8\textwidth]{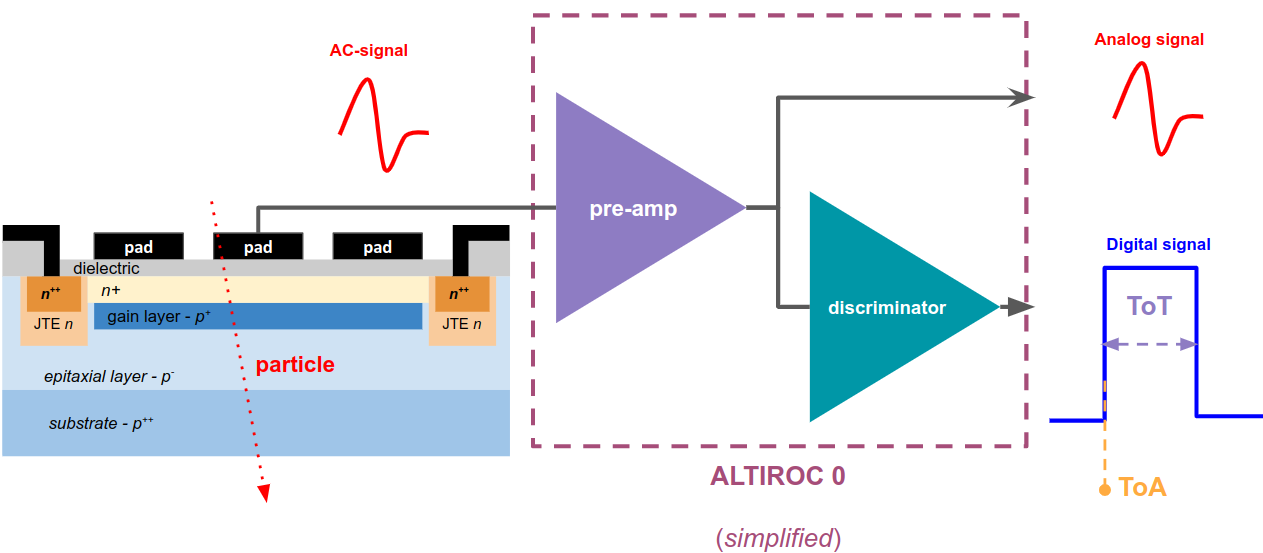}
    \caption{Simplified operation scheme of an AC-LGAD sensor read out by the ALTIROC 0 chip. The AC-signal generated by the interaction of a particle in silicon is multiplied in the voltage pre-amplifier stage, and then fed to a built-in discriminator to produce a digital signal. The width of such signal provides a measurement of the signal ToT.}
    \label{fig:simplifiedALTIscheme}
\end{figure}

The readout chip used in this study is the prototype ALTIROC 0 (version 2B), consisting of only the four voltage pre-amplifier and discriminator stages of the final ALTIROC chip. Two Time-to-Digital Converter (TDC) units can provide a digital signal encoding the information of the Time Of Arrival (ToA) and Time-Over-Threshold (TOT) of each input signal (Fig.~\ref{fig:simplifiedALTIscheme})~\cite{delataille:hal-02058308}. The rising edge of this digital pulse provides a measurement of the particle ToA while its width that of the ToT of the analog signal at the pre-amplifier stage~\cite{Agapopoulou_2020}.
The discriminator threshold is set by an external 10-bit DAC, common to all channels.
The ALTIROC 0 chip was developed to achieve a jitter smaller than 20~ps for an input charge of 10~fC, and provide a time measurement for input signals with charge as low as 4~fC. The ToT is used in the ALTIROC 0 design to correct for the signal time walk.
\subsection{Particle sources}
Measurements involving both an uncollimated flux of beta particles and a focused IR laser were performed to characterize the response to one or multiple mips of injected charge.
A 7.5~MBq Strontium-90 emitter was used as beta source and positioned on top of the AC-LGAD sensor, at a distance of approximately 2~cm. The source was enclosed in a 3D printed shell (made of Stratasys RGD875 polymer) with a single circular opening (6~mm radius) towards the sensor.
\begin{figure}[htbp]
    \centering
    \includegraphics[width=.75\textwidth]{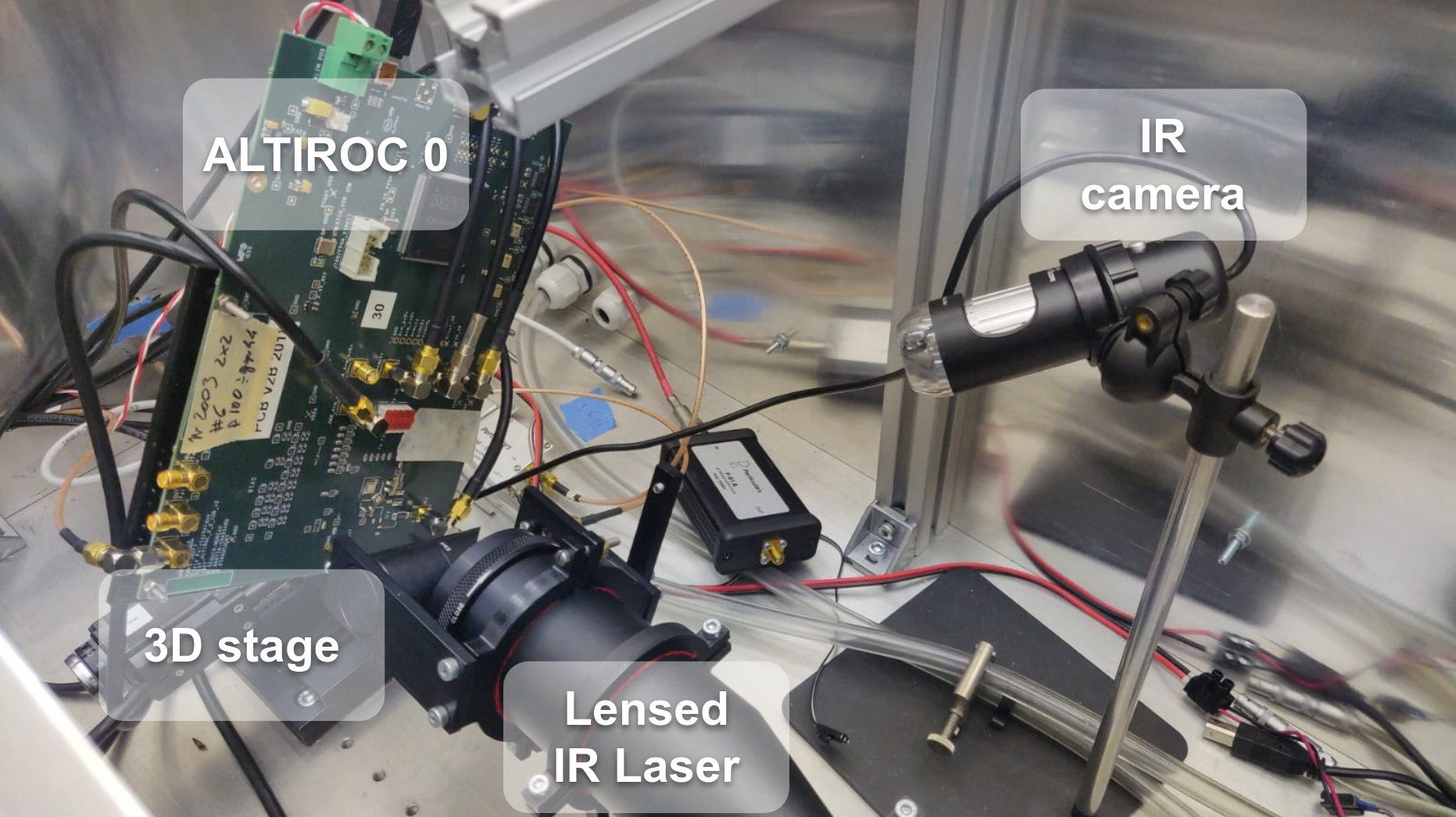}
    \caption{Picture of the TCT setup for the characterization of the AC-LGAD sensor using an IR laser. From left to right are shown the ALTIROC testing board mounted on a 3D stage with 1 $\mu$m precision; a lens for laser focusing; an IR camera used for alignment. The setup is enclosed in a light-tight aluminum box.}
    \label{fig:laser_position}
\end{figure}

A focused IR laser was obtained by means of a Particulars Scanning-Transient Current Technique (TCT) apparatus\footnote{http://www.particulars.si/products.php?prod=scanTCT.html}, shown in Fig.~\ref{fig:laser_position}. This setup was used to inject the equivalent of several mips of charge in the AC-LGAD sensor by means of an IR laser with a wavelength of 1064~nm. The ALTIROC 0 test-board was mounted on a 3-axis computer-controlled mechanical stage with a position resolution of less than 1 micron. The setup is enclosed in a light-tight aluminum box. Dedicated supports for the ALTIROC testing board were custom designed and 3D printed to improve the mechanical stability of the setup.
\section{Preamplifier response characterization}\label{sec:signal_characterization}
The sensor response was studied on all four connected strips using both the analog output of the ALTIROC VPA channels and the digital output generated by the on-board discriminators.

\subsection{Signal generated by a beta source (single mip)}
\label{sec:Measurements_Beta}
%
Betas originated by the decay of $^{90}$Sr and by its decay byproduct, $^{90}$Y, show a distribution of energy up to 2.2~MeV and can be considered in first approximation as mips. Both $^{90}$Sr and $^{90}$Y are almost pure $\beta^-$ emitter, with very little contamination from photon production. When interacting with silicon, the energy loss of betas is described by a Landau distribution. No focusing system has been applied to the betas.
\begin{figure}[htbp]
    \textit{a)}\hspace{.49\textwidth}\textit{b)}
    \centering
    \includegraphics[width=.49\textwidth, trim = 2 8 10 20, clip]{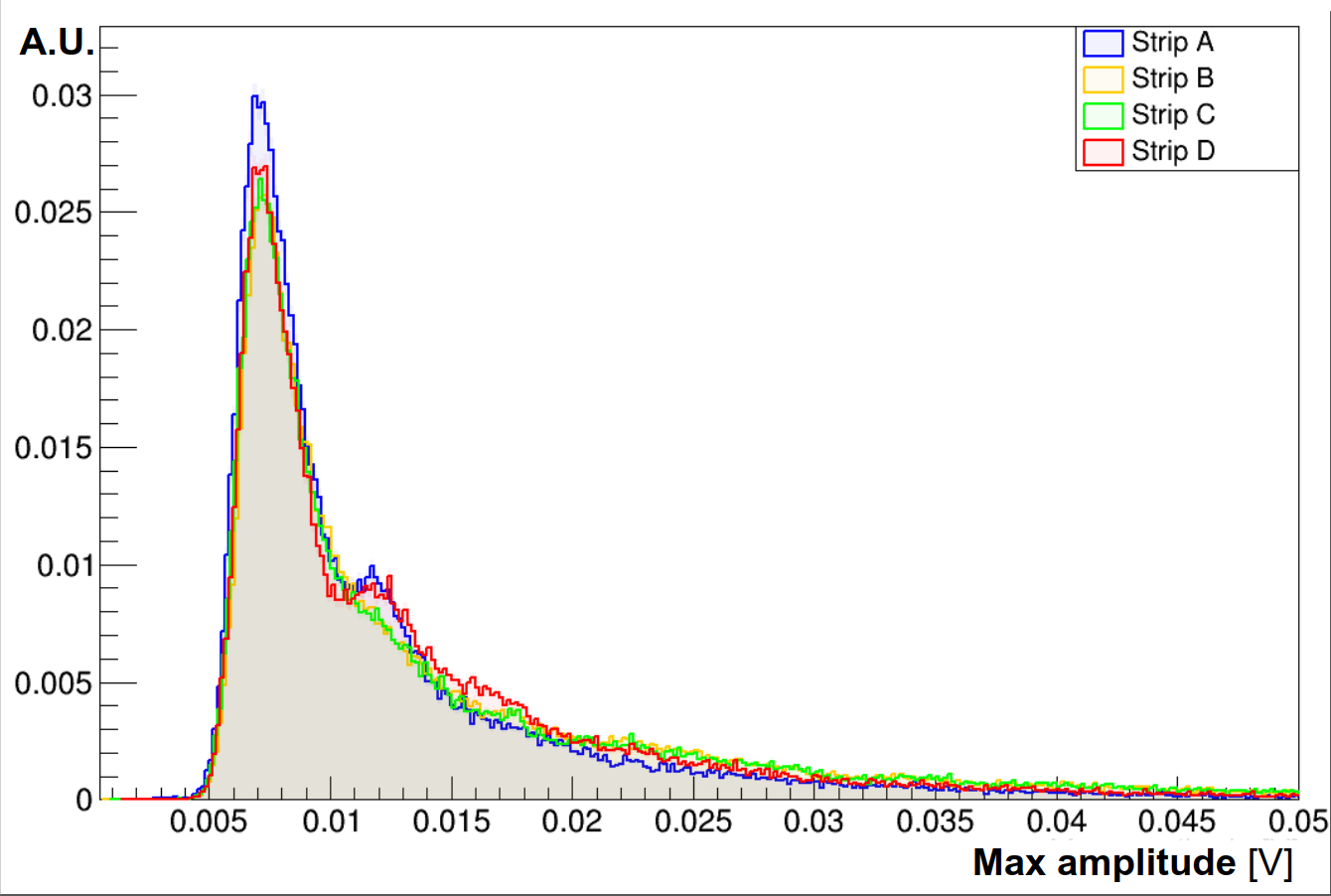}
    \includegraphics[width=.49\textwidth, trim = 2 8 10 20, clip]{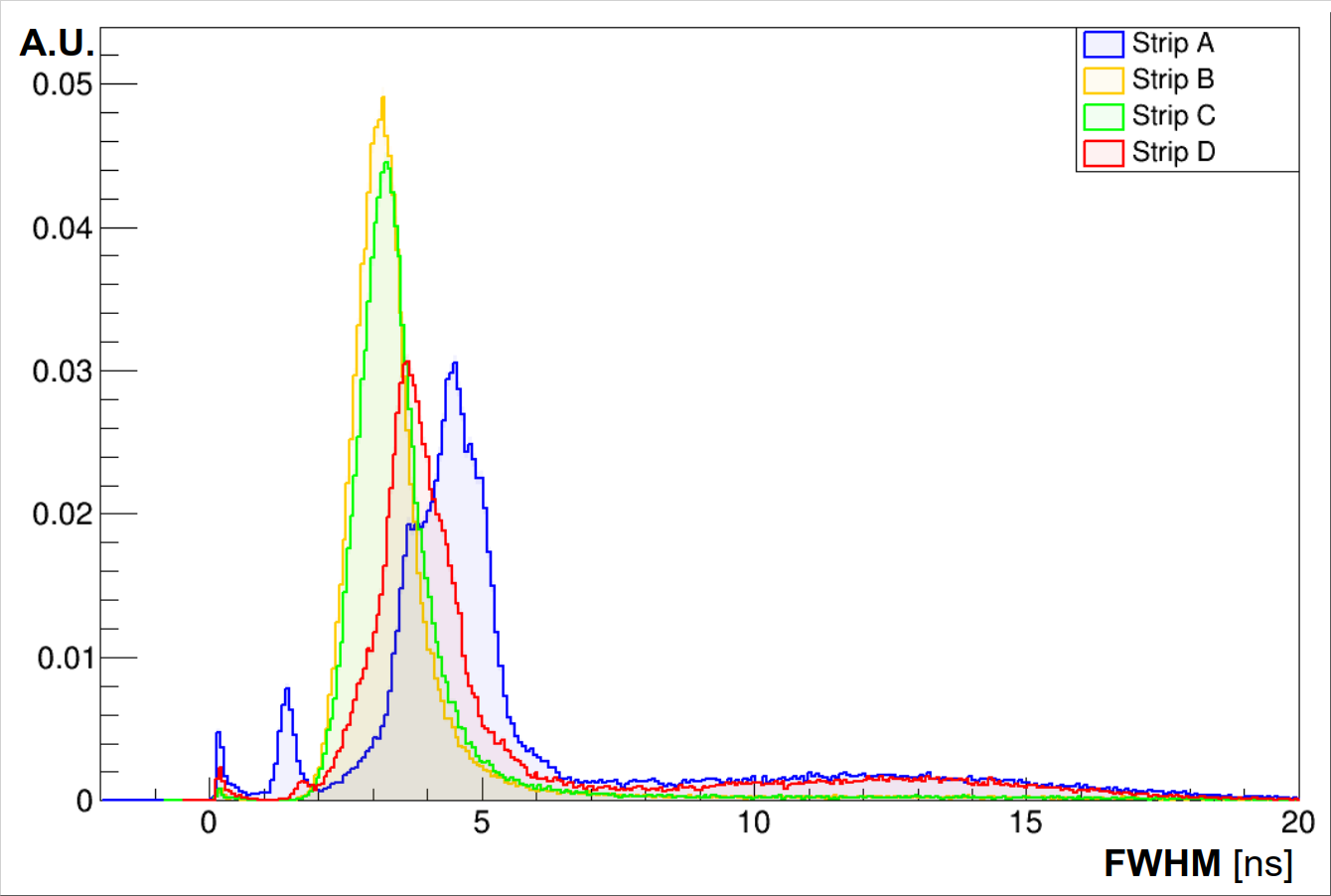}
    \caption{Distribution of the \textit{a}) signal amplitude  and \textit{b)} FWHM for beta signals interacting with the detector. The shapes of the two distributions for the central strips B and C are different from those observed for the lateral strips A and D.}
    \label{fig:beta_distribution}
\end{figure}
As shown in Fig.~\ref{fig:beta_distribution}, signals from strips B and C, close to the center of the device, have a FWHM of about 4~ns that is compatible to published results for standard (DC-)LGAD sensors readout via ALTIROC~\cite{Agapopoulou_2020}. Single mips signals have an average amplitude of $\sim$8~mV, with a long Landau tail up to amplitudes higher than 50~mV. The signal amplitude is computed as the maximum voltage (in absolute value) recorded by the oscilloscope above the baseline output voltage of the readout system.  
Lateral strips show slightly wider and more complex distributions, with multiple peaks in both FWHM and signal amplitudes distributions; this effect is hypothesised to be a signal sharing effect generated by the coupling of the lateral strips to the ground pad.
Since the system can read out only one strip at the time, and the beta beam is not collimated, it is not possible to disentangle signals from betas directly hitting the readout strip from signals generated by those hitting a neighbouring strip and detected in the readout strip due to signal sharing. A different setup proved to be necessary to study the properties of signal sharing.

\subsection{Laser response map}
A scanning TCT setup was used to evaluate the response of the detector at different impact positions of the incoming particle. Since the intensity of the laser can be tuned, it is possible to emulate the charge deposition equivalent to one or more mips on the sensor, while eliminating the statistical dispersion effects of the Landau distribution typical of charged particles thanks to the point-like energy release of photons in matter. 
\begin{figure}[htbp]
    \centering
    \includegraphics[width=.7\textwidth]{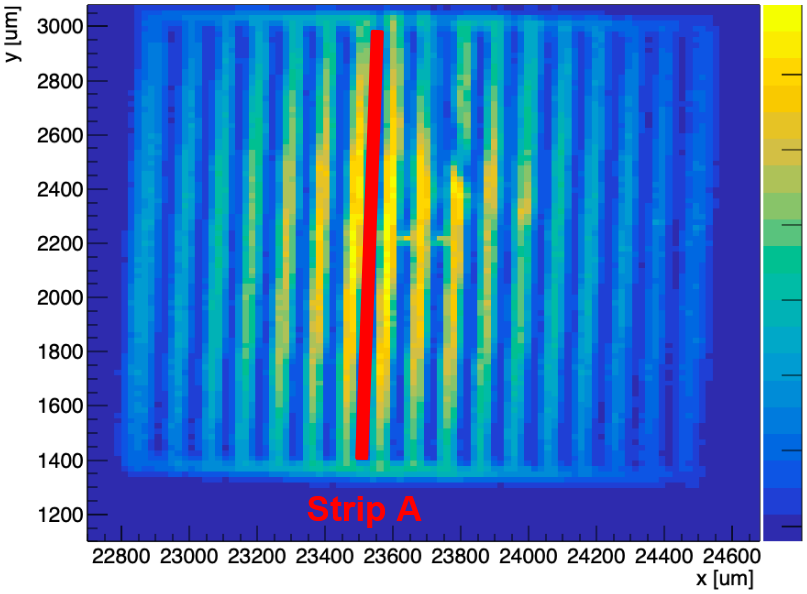}
    \caption{
    Map of charge deposited on Strip A as a function of the shining position of the IR laser, as read out by ALTIROC 0. Areas with higher collected charge are represented in yellow, while areas with lower collected charge are dark blue. Signal sharing can be visually identified by the yellow gradient around the readout strip A.}
    \label{fig:pixelmap_tct}
\end{figure}
Having under control the position of the incident photons, this setup allows us to estimate the signal characteristics and sharing inside the sensor. This is achieved by moving the sensor with respect to the laser focus-point and mapping the response of the AC-LGAD. The magnitude of the response is evaluated from the integral of the charge collected by the sensor as a function of the shining position of the IR laser. Fig.~\ref{fig:pixelmap_tct} shows the response map obtained from the output of Strip A as a function of the horizontal and vertical positions of the IR laser shining point.
Since the metallization layer of the electrodes blocks the IR laser, the sensor does not collect charge when the laser is shone on the strip metal.
The scan also allows to visualize the sharing of the signal in the AC-LGAD sensor: the readout strip senses charge even when the interaction occurs at a non-negligible distance from it. Signal can be  discriminated from noise at a distance of two pitches (200~$\mu$m) or more. As demonstrated in Ref.~\cite{Apresyan_2020} the signal from multiple strips can be used to precisely reconstruct the position of a charged particle on the sensor. The results in Fig.~\ref{fig:pixelmap_tct} show that signal sharing in AC-LGADs is preserved when the sensor are read out by an ALTIROC 0 chip. A complete characterization of the signal sharing is presented in Section \ref{subsec:signal_sharing}.
\subsection{Signals generated by an IR laser (multiple mips)}\label{subsec:signals_IR_alti}
Using the position information obtained with the pixel map on Fig.~\ref{fig:pixelmap_tct}, the laser was focused either between strips A and B (point L), between strips C and D (point R), or approximately at the center of the sensor, between strips B and C (point C), as detailed in Fig.\ref{fig:wirebonding_scheme}. A precise evaluation of the position of the laser focus-point on the sensor proved to be unachievable with the current setup, since vibrations in the mechanical stage due to the large board size caused shifts in the order of $\sim$~20~$\mu$m.
The IR laser deposits about 11~mips (32~mips) of equivalent charge onto the AC-LGAD sensor, estimated by comparing the observed signal amplitude to that computed for a single mip, when accounting for the readout impedance and the sensor gain.
\begin{figure}[htbp]
    \centering
    \textit{Left (L) point}\\\includegraphics[width=.45\textwidth,trim = 0 0 40 40,clip]{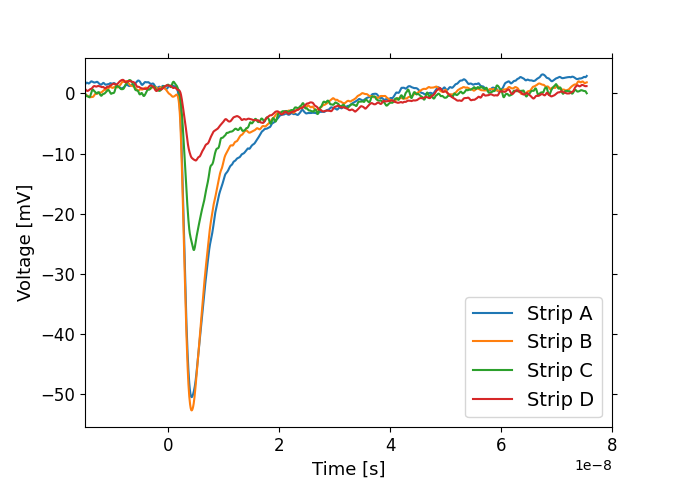}
    \includegraphics[width=.45\textwidth,trim = 0 0 40 40,clip]{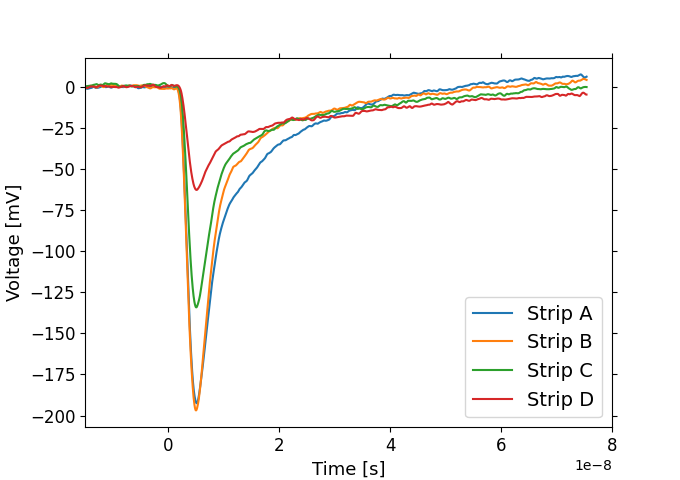}\\
    \textit{Center (C) point}\\\includegraphics[width=.45\textwidth,trim = 0 0 40 40,clip]{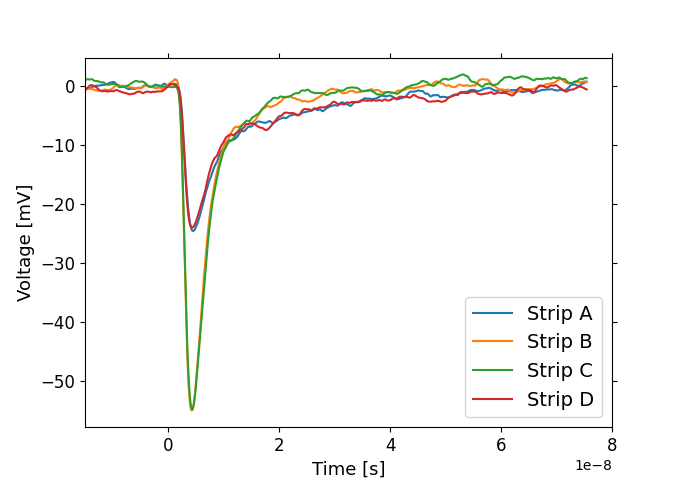}
    \includegraphics[width=.45\textwidth,trim = 0 0 40 40,clip]{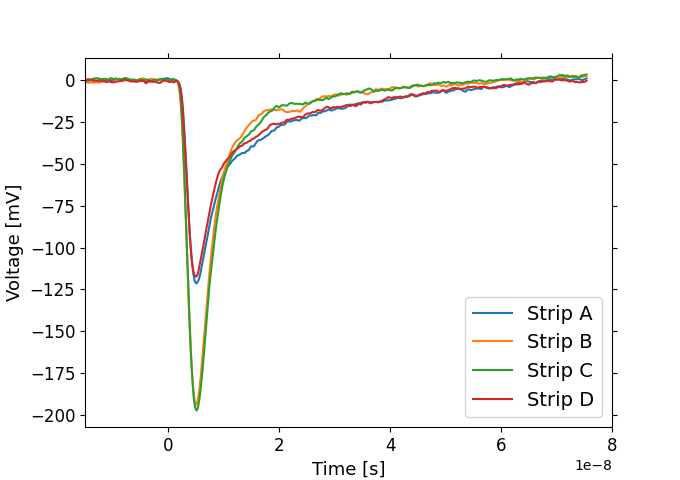}\\
    \textit{Right (R) point}\\\includegraphics[width=.45\textwidth,trim = 0 0 40 40,clip]{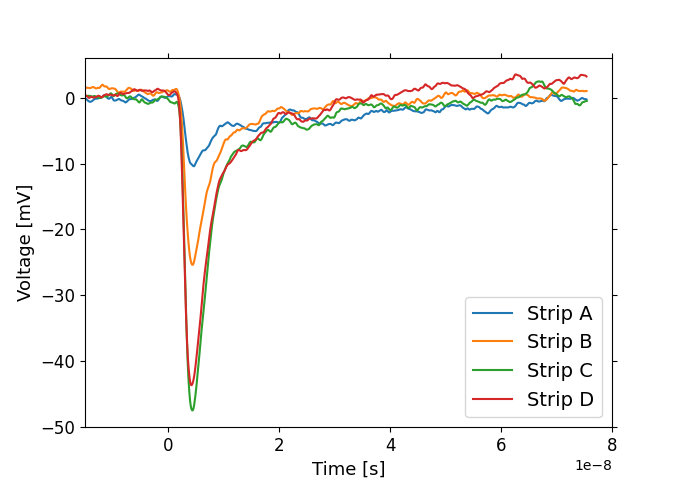}
    \includegraphics[width=.45\textwidth,trim = 0 0 40 40,clip]{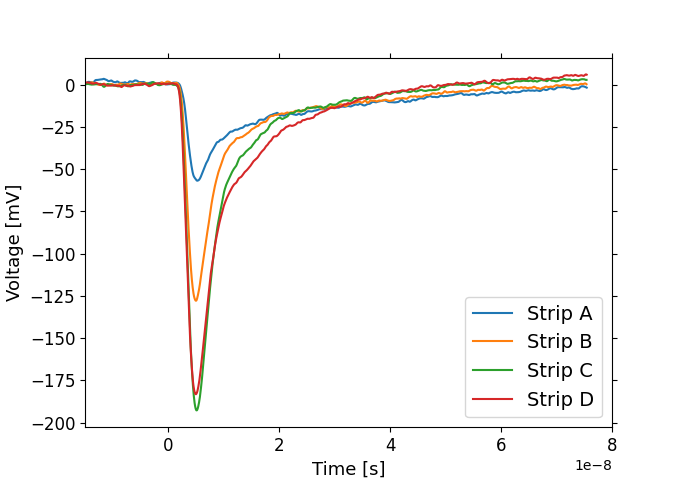}
    \caption{Examples of signals acquired at the VPA stage on four strips when the IR laser is focused at the left, center, or right  of the sensor. The laser injects 11 (\textit{left}) or 32 (\textit{right}) mips onto the sensor. The signal amplitude decreases linearly with distance due to the resistivity of the n$^{++}$ layer.}
    \label{fig:waveforms_tct}
\end{figure}

Examples of signals acquired at the VPA stage on the four strips when the IR laser is focused on each of the three tested positions (L, C and R) are shown in Fig.~\ref{fig:waveforms_tct}. 
When the laser deposits 11(32)~mips of charge onto the sensor it generates signals with an amplitude of $\sim$55~mV ($\sim$200~mV) on strips closer to the focus-point. The signal amplitude decreases linearly with the increase of distance due to the resistivity of the n$^{++}$ layer~ \cite{Tornago_2021}\cite{MANDURRINO2020163479}. This allows to estimate the linearity of the response of the detector and of the signal sharing when different charges are injected.
\begin{figure}[htbp]
    \centering
    \includegraphics[width=.8\textwidth]{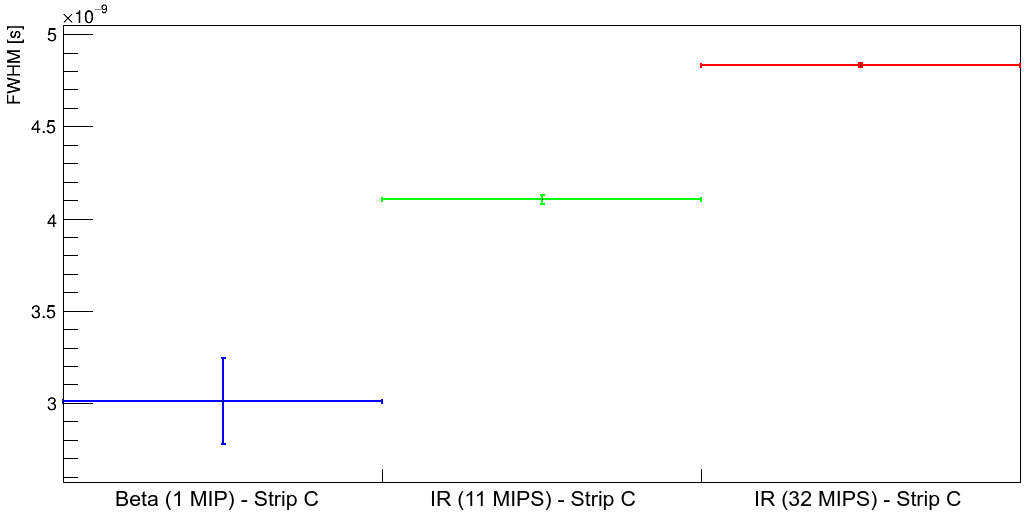}
    \caption{Most probable value of the analog signal FWHM for Strip C when the AC-LGAD sensor is irradiated with beta particles or an IR laser depositing 11 or 32~mips of charge. Value and error bands are extrapolated fitting each FWHM distribution using a Landau function. The FWHM value is compatible with published results for ALTIROC 0 using LGAD sensors and is dependant on the amount of mips of charge injected into the sensor.}
    \label{fig:altiroc_fwhm_comparison}
\end{figure}

Signals acquired using the VPA channels are fast, with a FWHM in the order of 4~ns for the strips near the laser point of incidence, compatible to what was observed for signals generated by beta particles in Sec.\ref{sec:Measurements_Beta}. The FWHM is dependant on the amount of mips of charge injected into the sensor, as seen in Fig.~\ref{fig:altiroc_fwhm_comparison}.

\subsection{Jitter measurement}\label{subsec:jitter}
\begin{figure}[htbp]
    \centering
    \includegraphics[width=.49\textwidth,trim = 30 20 0 0,clip]{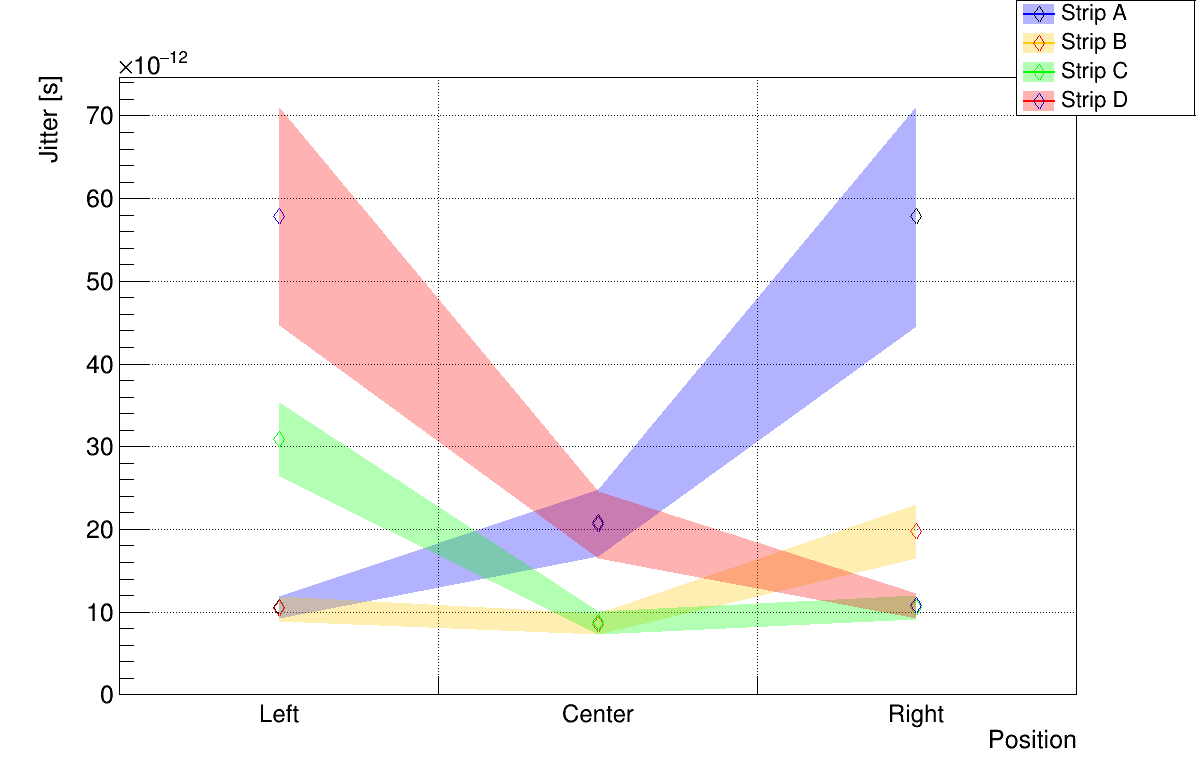}
    \includegraphics[width=.49\textwidth,trim = 30 20 0 0,clip]{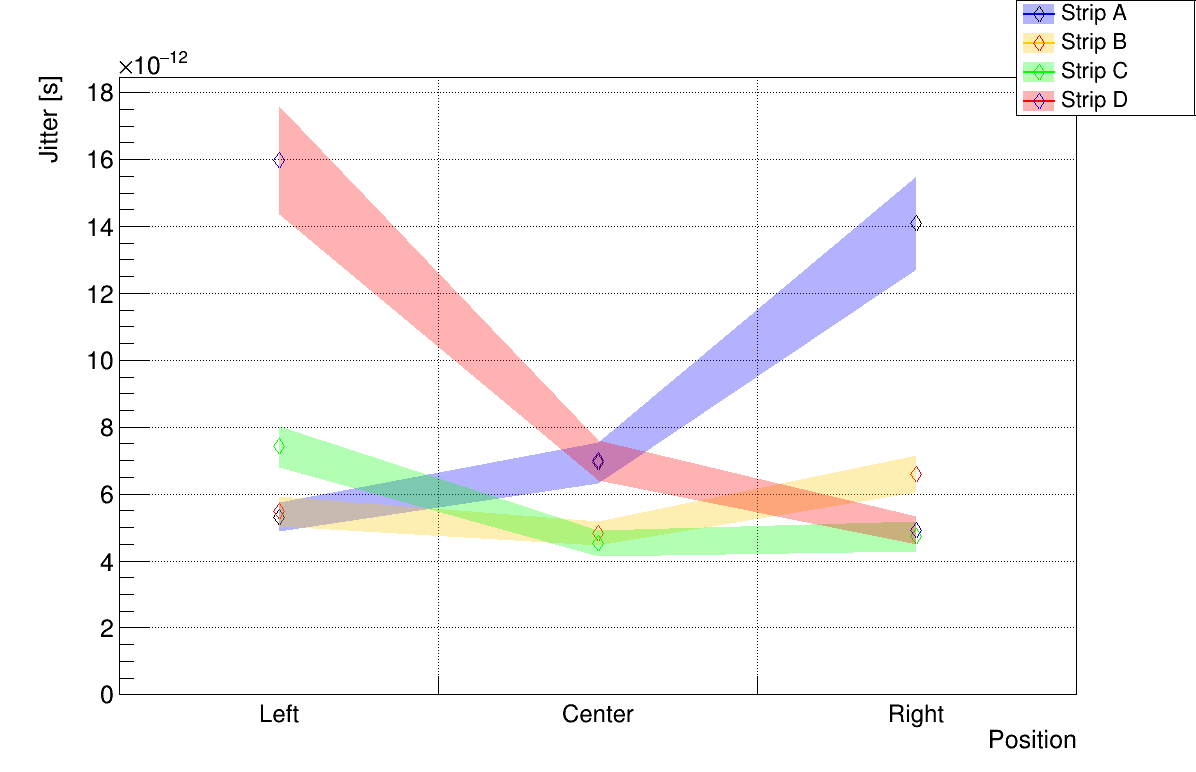}
    \caption{Jitter of analog signals from the ALTIROC 0 VPA channels when IR laser injects \textit{top}) 11 or \textit{bottom}) 32 mips, as a function of the laser shining position (Left, Center and Right). Signals from strips near the laser focus-point are fast, but the jitter degrades for farther strips due to the decrease in signal amplitude. The error bands represents the width of the Landau distribution obtained from data fitting.}
    \label{fig:tct_distribution_jitter}
\end{figure}

Signal jitter is a major contributor to the timing capabilities of any detector, and must therefore be characterized with precision. Contributions to signal jitter come from the readout electronic, from external conditions such as the environmental temperature, or are intrinsic to the detector. The total jitter is often computed as: 
\begin{equation}
   \textrm{jitter} = \sigma_{noise} \left( \frac{dV}{dt} \right)^{-1}
\end{equation}
where $\sigma_{noise}$ is the recorded signal noise and $dV/dt$ is the slew-rate of the signal computed between 30\% and 90\% of the signal maximum amplitude. The noise is estimated as the standard deviation of the voltage spread observed in the first 10$\%$ of all signals recorded by the oscilloscope. Measurements of signal jitter for all four strips are obtained from the analog VPA channels of ALTIROC 0 when injecting charge using the IR laser. Fig.~\ref{fig:tct_distribution_jitter} shows the jitter values between the four strips when the laser is shone onto the points L, C and R.
The mean value and overall shape of the computed jitter is consistent for all four analog channels, when accounting for the asymmetry generated by the setup mechanical oscillation previously described; signals generated by the 11~mips (32~mips) IR laser show a jitter of about 10~ps (6~ps) in the strips near the laser focus-point, compatible with previous observations and the ALTIROC 0 specifications for LGAD sensors. Strips farther from the laser focus-point show signals with a jitter of 25 to 35~ps (7 to 9~ps) at one pitch of distance and 66-67~ps (17 to 19~ps) at two pitches of distance. This degradation of the signal jitter is a direct effect of the observed decrease of signal amplitude on strips far from the laser focus-point. 

\subsection{Comparison to DRF board}\label{subsec:signals_IR_fnal}

Results obtained using the aforementioned detector were compared to those measured using a similar system based on a Discrete Radio Frequency (DRF) amplifier board designed and produced by FermiLab. Such board is widely used for prototyping and testing of LGAD and AC-LGAD sensors and is considered a good benchmark for fast-time multi-channel acquisition of silicon sensor signals. This DRF board, mounting an AC-LGAD sensor identical to the one tested together with the ALTIROC 0, was installed on the scanning TCT system. All sixteen strips were connected to the board and four central strips, labelled again as strips A, B, C, D, are selected to be on the same respective positions as the those wire-bonded in the ALTIROC 0 chip to allow for a direct comparison between the two setups as detailed in Fig.~\ref{fig:wirebonding_scheme}.
\begin{figure}[htbp]
    \centering
    \textit{Left (L) point}\\
    \includegraphics[width=.49\textwidth,trim = 0 0 30 40,clip]{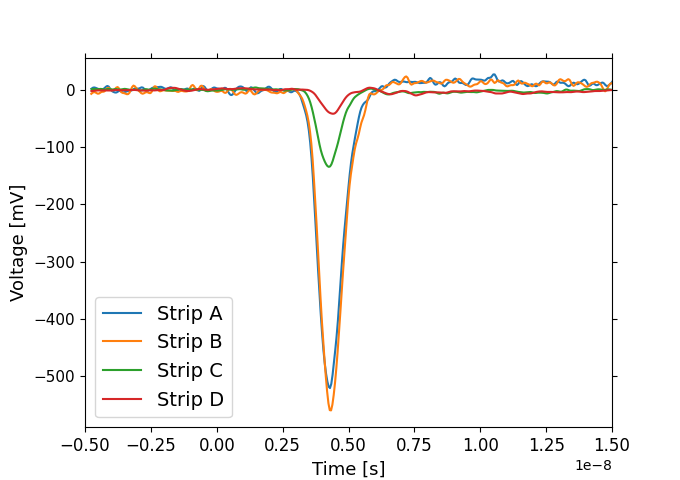}
    \includegraphics[width=.49\textwidth,trim = 0 0 30 40,clip]{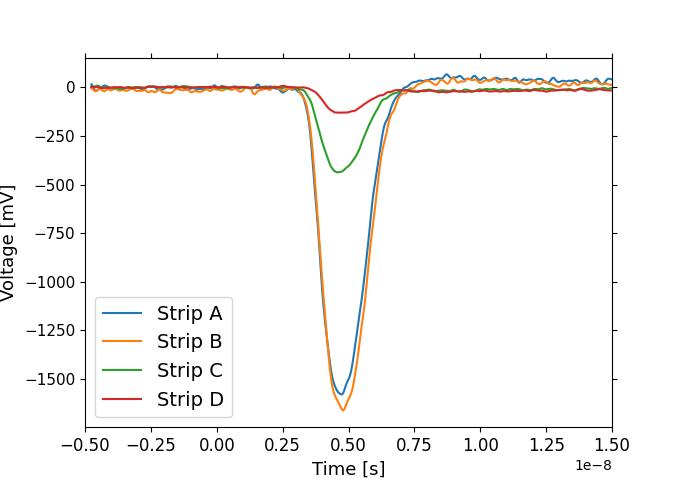}\\
    \textit{Center (C) point}\\
    \includegraphics[width=.49\textwidth,trim = 0 0 30 40,clip]{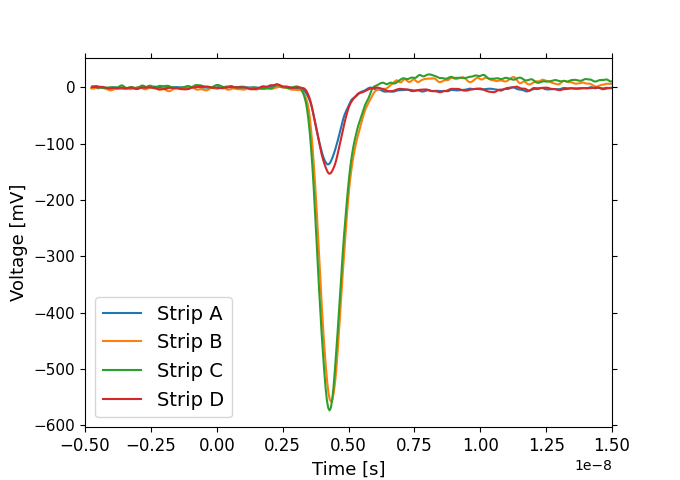}
    \includegraphics[width=.49\textwidth,trim = 0 0 30 40,clip]{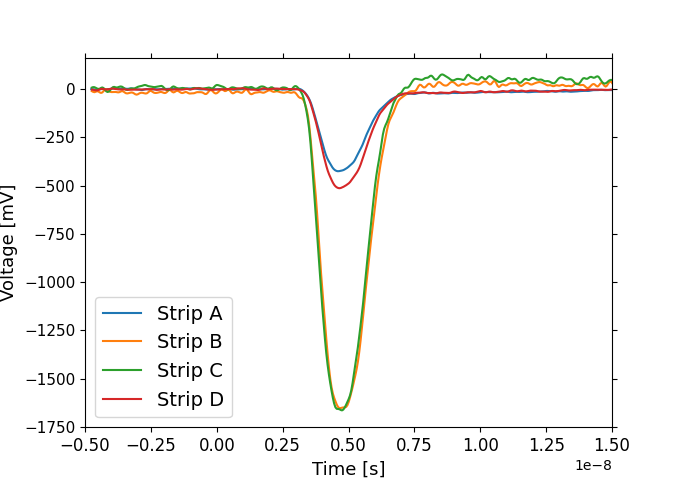}\\
    \textit{Right (R) point}\\
    \includegraphics[width=.49\textwidth,trim = 0 0 30 40,clip]{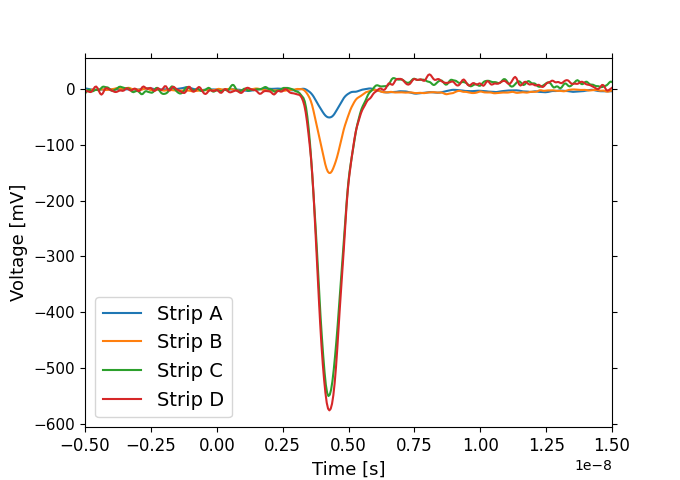}
    \includegraphics[width=.49\textwidth,trim = 0 0 30 40,clip]{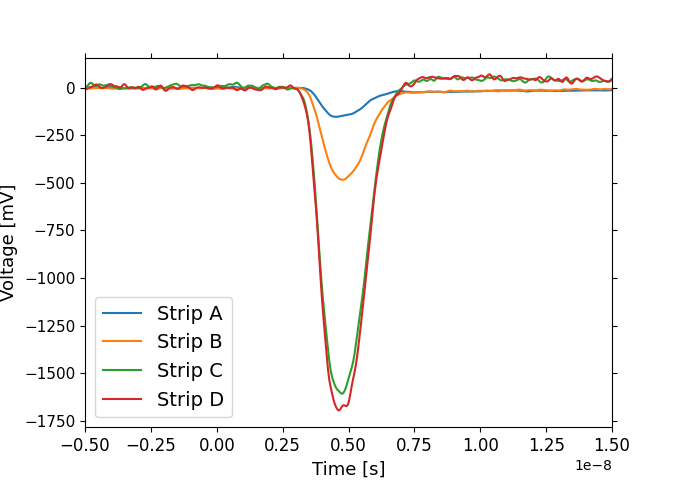}
    \caption{Examples of signals on four strips connected to the DRF board when the IR laser is focused at the left, center, or right of the sensor. The laser injected 11 (\textit{left}) or 32 (\textit{right}) mips onto the sensor.}
    \label{fig:waveforms_tct_fnal}
\end{figure}
Examples of signals acquired from the four strips when the IR laser is focused on each of the three tested positions L, C and R are shown in Fig.~\ref{fig:waveforms_tct_fnal}. 
When the laser deposits 11(32)~mips of charge onto the sensor it generates signals with an amplitude of $\sim$590~mV ($\sim$1700~mV) on strips closer to the focus-point and, as in the previous case, the signal amplitude decreases linearly with the increase of distance.
The amplitudes of signals acquired from the DRF board are significantly (by a factor $\sim$8.5) larger than those readout by the ALTIROC 0 setup using identical sensors and in the same conditions, due to the different gain of the two readout systems.
\subsection{Signal sharing}\label{subsec:signal_sharing}
The signal sharing is evaluated by measuring the amplitudes of signals generated by the IR laser on four neighbouring strips. The percentage of shared signal is then computed as the ratio between the signal amplitude measured by each strip and that measured by the strip closest to the focus-point.
\begin{figure}[htbp]
    \centering
    \includegraphics[width=\textwidth,trim = 80 0 80 0,clip]{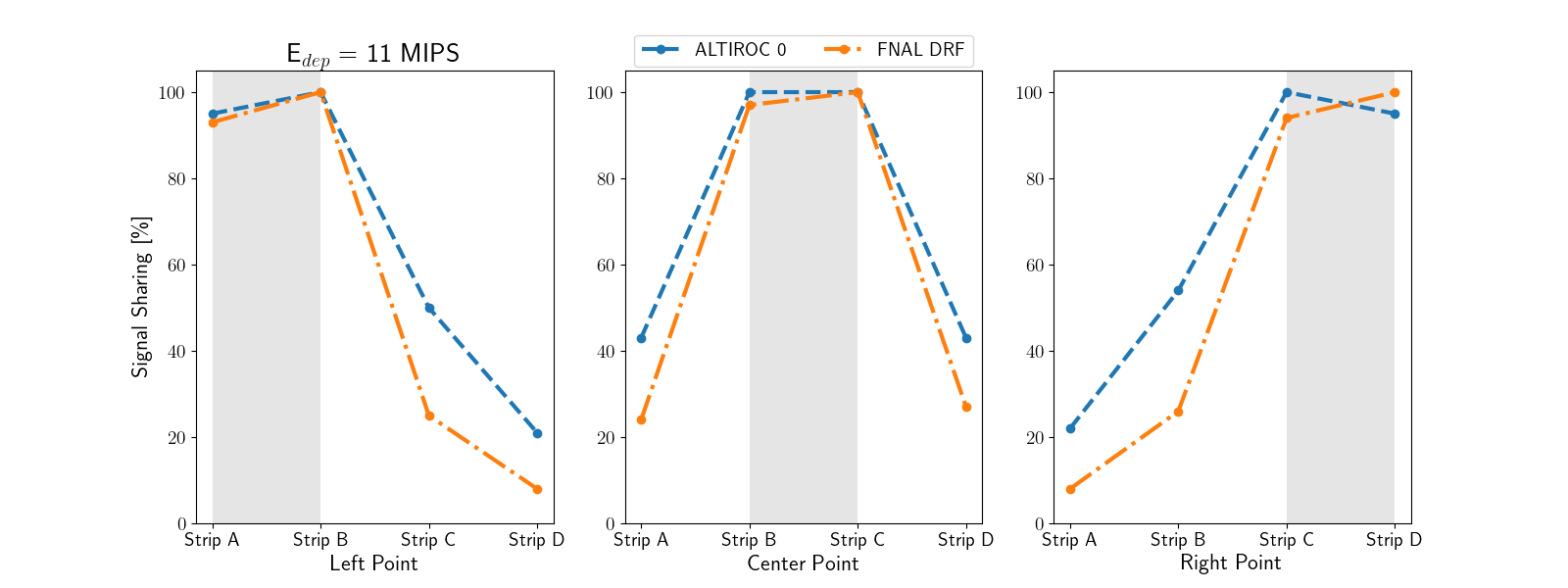}
    \includegraphics[width=\textwidth,trim = 60 0 80 0,clip]{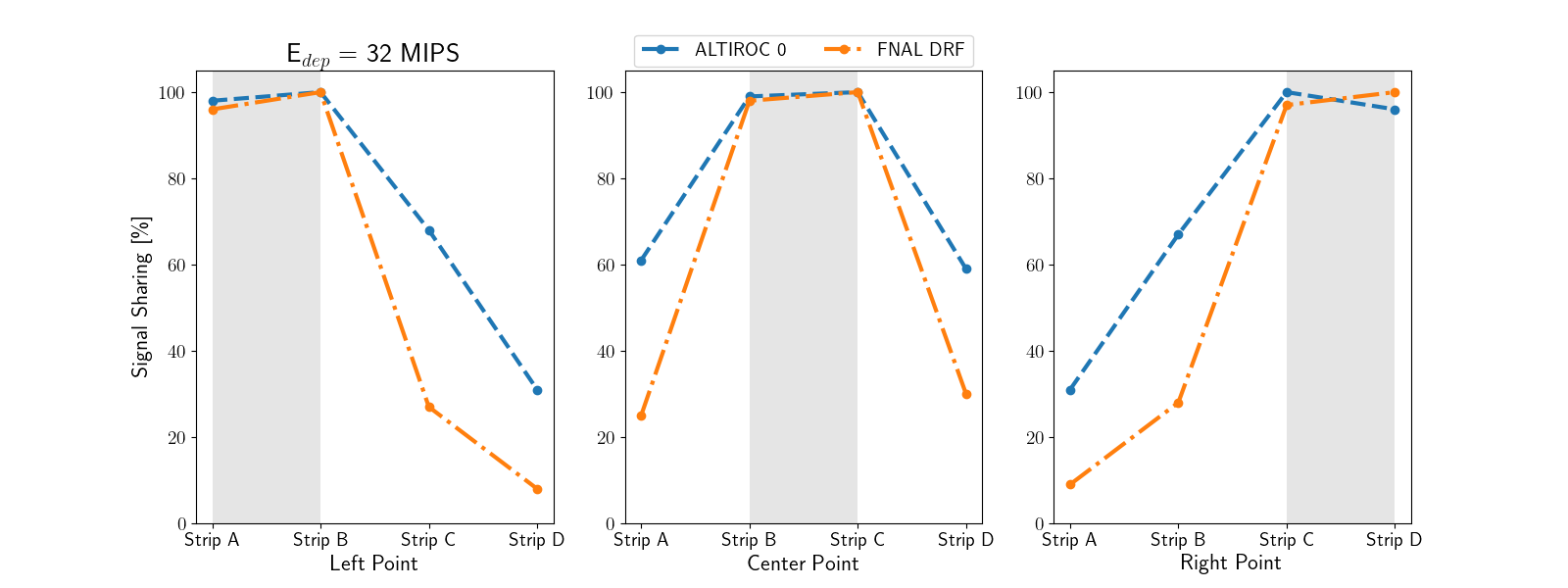}
    \caption{Sharing profile on AC-LGAD strips read out by the ALTIROC 0 chip and the Fermilab (FNAL) DRF board at different distances from the focus-point of an IR laser. The laser injects 11~mips (\textit{top}) or 32~mips (\textit{bottom}) on the sensor. The gray area in each window identifies the focus-position of the laser.}
    \label{fig:signal_sharing}
\end{figure}
The evaluation of the signal sharing profile on AC-LGAD has been measured for the sensors mounted on both the ALTIROC 0 chip and the DRF readout board using the data presented in Sec.~\ref{subsec:signals_IR_alti} and \ref{subsec:signals_IR_fnal}.
Fig.~\ref{fig:signal_sharing} details the signal sharing profile induced on the AC-LGADs as a function of the laser shining position when injecting either 11 or 32 mips. The sharing profiles are symmetrical with respect to the sensor central axis within uncertainty. In the case of the ALTIROC 0 setup, a small decrease of signal sharing is observed when the laser is shone at the center point compared to that observed in left and right points at the same distance; this is likely a border effect caused by the presence of several non-bonded strips on the sensor. This is a small effect in magnitude. 
It can be also observed that signal sharing is 2-3 times more prominent using the DRF setup when accounting for the strip position and laser intensity. This difference in signal sharing profile in the two readouts is expected due to the different trans-impedance and capacitance of ALTIROC 0 with respect to the DRF amplifiers. 
\section{Discriminator response characterization}
\label{sec:digital_characterization}

The ability of the ALTIROC 0 TDCs to encode and transmit the signal ToT is important for the construction of an AC-LGAD based detector: when combined with the knowledge of the signal sharing profile of the detector, it allows to precisely reconstruct the position of the incoming particle at a sub-pitch level~\cite{Apresyan_2020}, as it approximates the value of the deposited charge. To this end, the chip ToT response has to be mapped univocally to the signal shared on each strip of the sensor.
\begin{figure}[htbp]
    \centering
    \includegraphics[width=.49\textwidth,trim = 0 0 40 40,clip]{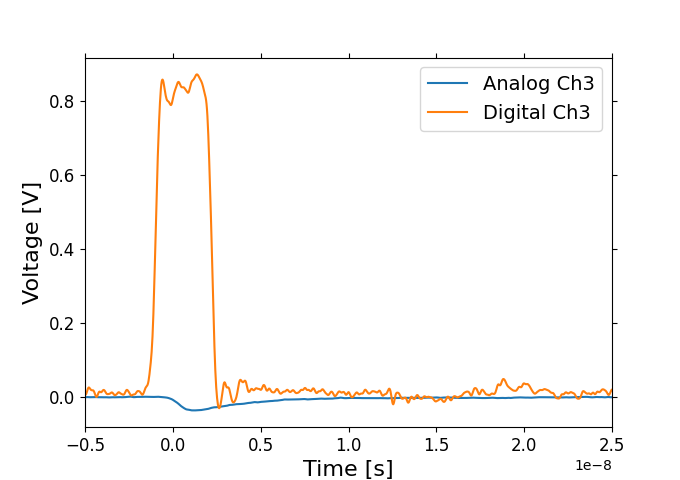}
    \includegraphics[width=.49\textwidth,trim = 0 0 40 40,clip]{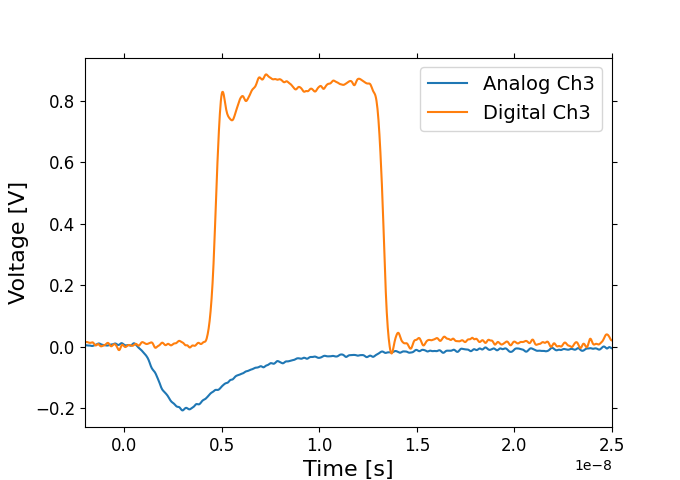}
    \caption{Examples of analog and digital output signals of ALTIROC 0 extracted from Strip C, and generated by the interactions of betas (\textit{left}) and IR laser (\textit{right}). The width of the digital signal, encoding the ToT, is proportional to the amplitude of the analog signal.}
    \label{fig:waveforms_tct_A+D}
\end{figure}
Figure~\ref{fig:waveforms_tct_A+D} shows the digital response of the ALTIROC 0 discriminator compared to the respective analog pre-amplifier output when the sensor is irradiated with either betas or an IR laser. The width of this digital signal encodes the ToT of the analog signal generated in the pre-amplifier stage.

\begin{figure}[htbp]
    \centering
    \includegraphics[width=.8\textwidth,trim = 30 20 10 50,clip]{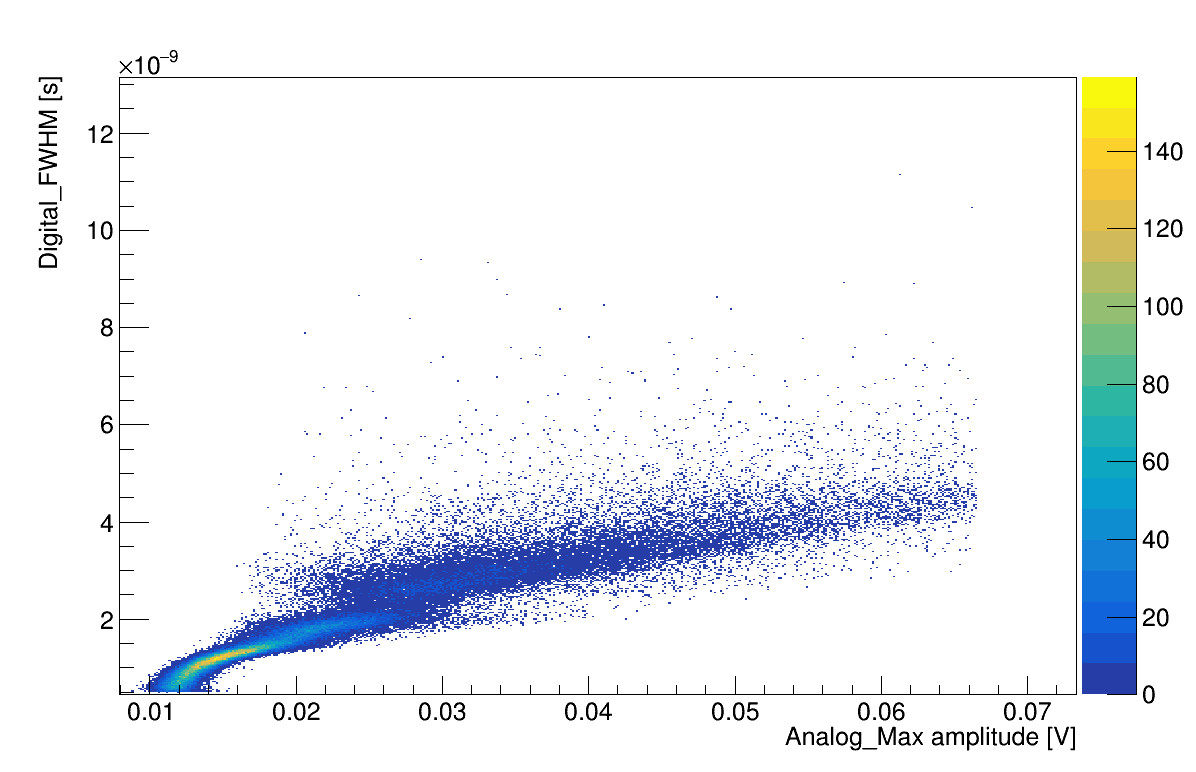}
    \caption{FWHM of the discriminator output of the ALTIROC 0 ASIC as a function of the respective amplitude of the analog signals, for signals generated by $\beta$ particles. The FWHM of the digital signal shows a clear univocal dependence on the analog signal amplitude and therefore on the deposited energy.}
    \label{fig:width_vs_amplitude}
\end{figure}

A characterization of the discriminator response can be obtained by correlating the FWHM of the discriminator digital output to the ToT of the analog signal. Signals from beta interactions with a wide range of deposited charge allows to evaluate this response characteristic, as shown in Fig.~\ref{fig:width_vs_amplitude}. Since the amplitude and the ToT of the analog signal are strictly correlated, the former is used as proxy of the latter for this evaluation.

Since the threshold of the discriminator can be adjusted, different thresholds have been applied in the two cases to compensate for the different charge deposited by the betas and the IR laser.
The effect of Landau fluctuations in interactions with betas produces a long tail in the distribution of analog signal amplitude. This is reflected in the measurement of the ToT, where a clear univocal dependence of the digital signal width to the analog signal amplitude can be observed.

\section{Conclusions}
\label{sec:Conclusions}
The first characterization of AC-LGAD signals read out by an ASIC is presented using the ALTIROC 0 readout ASIC. While designed to acquire DC-coupled signals from LGAD sensors, ALTIROC 0 proved to be a suitable readout for bipolar AC-coupled signals generated in BNL-made AC-LGADs. The output of the detector formed by an AC-LGAD sensor coupled with the ALTIROC 0 chip was studied by injecting one to multiple mips of charge on the sensor using either beta particles or a focused IR laser. Characteristics of the readout signals such as amplitude, shape and FWHM are compatible with those measured with test-boards used in test-beams to read out AC-LGADs and with previous ALTIROC results obtained for standard, i.e. DC-coupled, LGADs.

The linearity of the ToT measurement provided by the ALTIROC 0 discriminator over a wide range of injected charges allows to exploit signal sharing and in turn improve space resolution in future experiments beyond (pitch size)/$\sqrt{12}$ as in typical discrete detectors that use a binary readout system. Such experiments may include high granularity detectors for high-energy and nuclear physics applications, e.g. at the EIC. It has been proven that the signal sharing profile of the sensor is affected by the trans-impedance and capacitance of the readout chip; these parameters can be fine tuned together with the sensor pitch to provide the targeted spatial resolution while keeping the sensor segmentation sufficiently sparse to minimise the number of readout channel, and in turn readout bandwidth, cooling and costs.


These studies can be used as a stepping stone for the development of novel types of readout chips, based on the ALTIROC technology and tailored to match AC-LGAD signal properties, aimed at 4D reconstruction of particles by exploiting their signal sharing capabilities.

\section*{Funding}
This material is based upon work supported by the U.S. Department of Energy under grant DE-SC0012704. This research used resources of the Center for Functional Nanomaterials, which is a U.S. DOE Office of Science Facility, at Brookhaven National Laboratory under Contract No. DE-SC0012704.

\section*{Acknowledgments}
The authors  wish to thank their colleagues at Brookhaven National Laboratory: Ron Angona and  Sean Robinson for sensor fabrication; Don Pinelli, Joe Pinz, Antonio Verderosa and Tim Kersten for board assembly and wire bonding. 

\bibliographystyle{unsrt}
\bibliography{main}

\begin{thebibliography}{10}

\bibitem{Fern_ndez_Mart_nez_2016}
P.~Fern{\'{a}}ndez-Mart{\'{\i}}nez, D.~Flores, and S.~Hidalgo et~al.
\newblock Design and fabrication of an optimum peripheral region for low gain
  avalanche detectors.
\newblock {\em Nuclear Instruments and Methods in Physics Research Section A:
  Accelerators, Spectrometers, Detectors and Associated Equipment},
  821:93--100, jun 2016.

\bibitem{Cartiglia_2017}
N.~Cartiglia, A.~Staiano, and V.~Sola et~al.
\newblock Beam test results of a 16 ps timing system based on ultra-fast
  silicon detectors.
\newblock {\em Nuclear Instruments and Methods in Physics Research Section A:
  Accelerators, Spectrometers, Detectors and Associated Equipment}, 850:83--88,
  apr 2017.

\bibitem{HGTD}
{Technical Proposal: A High-Granularity Timing Detector for the ATLAS Phase-II
  Upgrade}.

\bibitem{MTD}
Joel~N. Butler and Tommaso Tabarelli~de Fatis.
\newblock {A MIP Timing Detector for the CMS Phase-2 Upgrade}.
\newblock 2019.

\bibitem{Andra:ay5534}
M.~Andr{\"{a}}, J.~Zhang, and A.~Bergamaschi et~al.
\newblock {Development of low-energy X-ray detectors using LGAD sensors}.
\newblock {\em Journal of Synchrotron Radiation}, 26(4):1226--1237, Jul 2019.

\bibitem{aclgad_1}
N.~Cartiglia.
\newblock Issues in the design of ultrafast silicon detectors.
\newblock {\em TREDI2015, Trento, Italy}.

\bibitem{aclgad_2}
RSD Collaboration.
\newblock {RSD} - resistive {AC}-coupled silicon detectors.
\newblock {\em
  https://www.to.infn.it/attivita-scientifica/ricerca-tecnologica/RSD/}.

\bibitem{aclgad_3}
RD50 Collaboration.
\newblock {RD50} - radiation hard semiconductor devices for very high
  luminosity colliders.
\newblock {\em http://rd50.web.cern.ch/rd50/}.

\bibitem{aclgad_4}
AIDA-2020 Collaboration.
\newblock {AIDA-2020} - advanced european infrastructures for detectors at
  accelerators, {WP7:} advanced hybrid pixel detectors.

\bibitem{Apresyan_2020}
A.~Apresyan, W.~Chen, and G.~D’Amen et~al.
\newblock Measurements of an {AC-LGAD} strip sensor with a 120 {GeV} proton
  beam.
\newblock {\em Journal of Instrumentation}, 15(09):P09038–P09038, Sep 2020.

\bibitem{Heller_2022}
R.~Heller, C.~Madrid, A.~Apresyan, W.K. Brooks, W.~Chen,
  G.~D{\textquotesingle}Amen, G.~Giacomini, I.~Goya, K.~Hara, S.~Kita, S.~Los,
  A.~Molnar, K.~Nakamura, C.~Pe{\~{n}}a, C.~San Mart{\'{\i}}n, A.~Tricoli,
  T.~Ueda, and S.~Xie.
\newblock Characterization of {BNL} and {HPK} {AC}-{LGAD} sensors with a 120
  {GeV} proton beam.
\newblock {\em Journal of Instrumentation}, 17(05):P05001, may 2022.

\bibitem{Giacomini_2019_LGAD}
G.~Giacomini, W.~Chen, and F.~Lanni et~al.
\newblock Development of a technology for the fabrication of low-gain avalanche
  diodes at {BNL}.
\newblock {\em Nuclear Instruments and Methods in Physics Research Section A:
  Accelerators, Spectrometers, Detectors and Associated Equipment},
  934:52–57, Aug 2019.

\bibitem{Giacomini_2019_ACLGAD}
G.~Giacomini, W.~Chen, and G.~D’Amen et~al.
\newblock Fabrication and performance of {AC-coupled LGADs}.
\newblock {\em Journal of Instrumentation}, 14(09):P09004–P09004, Sep 2019.

\bibitem{romanpots}
E.C. Aschenauer and A.Tricoli.
\newblock {EIC RD} progress report {eRD24}.

\bibitem{delataille:hal-02058308}
C.~de~La~Taille, S.~Callier, and S.C. Di~Lorenzo et~al.
\newblock {ALTIROC0, a 20 pico-second time resolution ASIC for the ATLAS High
  Granularity Timing Detector (HGTD)}.
\newblock In {\em {Topical Workshop on Electronics for Particle Physics}},
  volume TWEPP-17, page 006, Santa Cruz, United States, September 2017.

\bibitem{Agapopoulou_2020}
C.~Agapopoulou, S.~Blin, and A.~Blot et~al.
\newblock {Performance of a Front End prototype ASIC for picosecond precision
  time measurements with LGAD sensors}.
\newblock {\em Journal of Instrumentation}, 15(07):P07007–P07007, Jul 2020.

\bibitem{Tornago_2021}
M.~Tornago, R.~Arcidiacono, and N.~Cartiglia et~al.
\newblock Resistive {AC}-coupled silicon detectors: Principles of operation and
  first results from a combined analysis of beam test and laser data.
\newblock {\em Nuclear Instruments and Methods in Physics Research Section A:
  Accelerators, Spectrometers, Detectors and Associated Equipment},
  1003:165319, jul 2021.

\bibitem{MANDURRINO2020163479}
M.~Mandurrino, R.~Arcidiacono, and M.~Boscardin et~al.
\newblock {Analysis and numerical design of Resistive AC-Coupled Silicon
  Detectors (RSD) for 4D particle tracking}.
\newblock {\em Nuclear Instruments and Methods in Physics Research Section A:
  Accelerators, Spectrometers, Detectors and Associated Equipment}, 959:163479,
  2020.

\end{thebibliography}
\end{document}